# Treedy: A Heuristic for Counting and Sampling Subsets

**Teppo Niinimäki**
Helsinki Institute for Information Technology
Deparment of Computer Science
University of Helsinki
teppo.niinimaki@cs.helsinki.fi

**Mikko Koivisto**
Helsinki Institute for Information Technology
Deparment of Computer Science
University of Helsinki
mikko.koivisto@cs.helsinki.fi

## Abstract

Consider a collection of weighted subsets of a ground set $N$. Given a query subset $Q$ of $N$, how fast can one (1) find the weighted sum over all subsets of $Q$, and (2) sample a subset of $Q$ proportionally to the weights? We present a tree-based greedy heuristic, Treedy, that for a given positive tolerance $d$ answers such counting and sampling queries to within a guaranteed relative error $d$ and total variation distance $d$, respectively. Experimental results on artificial instances and in application to Bayesian structure discovery in Bayesian networks show that approximations yield dramatic savings in running time compared to exact computation, and that Treedy typically outperforms a previously proposed sorting-based heuristic.

## 1 INTRODUCTION

Reasoning with probabilistic models typically deals with queries on sets of weighted points. For instance, one may ask a maximum-weight point subject to the constraint that the point satisfies some given property. Likewise, one may ask the total weight (e.g., the probability mass) of the points or a random point sampled proportionally to the weights, subject to the given constraint. The large number of points and the complexity of the constraint often render these tasks computationally very challenging.

The present works addresses a particular class of queries, we call *subset queries*, formalized as follows. Let $N$ be a ground set of $n$ elements and $\mathcal{C}$ a collection of $m$ subsets of $N$, each subset $S \in \mathcal{C}$ associated with a non-negative weight $w(S)$. For convenience, we extend the weight function to all subsets of the ground set by letting $w(S) = 0$ for $S \notin \mathcal{C}$. A *subset counting query* asks the total weight of all subsets of a given query subset $Q \subseteq N$, given by

$$W(Q) = \sum_{S \subseteq Q} w(S). \tag{1}$$

Analogously, a *subset sampling query* asks a random subset of the query subset $Q$, such that any particular subset $S \subseteq Q$ gets selected with probability $w(S)/W(Q)$. While queries about maxima, minima, median, etc. can be defined in a similar fashion, we will restrict ourselves to counting and sampling queries in the sequel.

Our focus will be on scenarios where $m$, the size of the collection, is much smaller than $2^n$, the number of all subsets of the ground set. The opposite case when $m$ is close to $2^n$ is also fundamental but, to a large extent, well understood. Namely, using the fast zeta transform algorithm (Yates, 1937; Kennes and Smets, 1990; Kennes, 1991; Koivisto and Sood, 2004; Koivisto, 2006) the values $W(Q)$ can be computed for all $Q \subseteq N$ in a total of roughly $n2^n$ additions, after which answering any counting or sampling query is very fast.

### 1.1 APPLICATION: ORDER-MCMC FOR BAYESIAN NETWORK LEARNING

A motivating example of subset queries is provided by the order-MCMC method of Friedman and Koller (2003) for learning the directed acyclic graph (DAG) of a Bayesian network model (Pearl, 1988, 2000; Buntine, 1991; Heckerman et al., 1995). For closely related recent developments that likewise involve subset queries, see the works of Ellis and Wong (2008), Niinimäki et al. (2011), and Niinimäki and Koivisto (2013).

Order-MCMC samples node orderings by simulating a Markov chain whose stationary distribution is the posterior distribution. The time consumption of order-MCMC is determined by the complexity of evaluating the posterior probability (up to a normalizing constant) of a given node ordering $v_1 v_2 \cdots v_n$. Ef-ficient evaluation is facilitated by the factorization of the posterior into a product $W_1 W_2 \cdots W_n$, where each factor $W_j$ is given by (1) as the total weight $W(\{v_1, v_2, \ldots, v_{j-1}\})$ for a weight function $w$ that depends on the node $v_j$. The interpretation is that any subset of the nodes preceding $v_j$ can form the set of *parents* of $v_j$ in the network, the weight indicating how well a particular selection of parents fits the data and the prior beliefs.

To quantify the time requirements in this example, suppose that each node is allowed to have at most $k$ parents—having a relatively small $k$ is a common practice unless $n$ is very small. Then each $W_j$ can be evaluated using about $\binom{j-1}{k}_* = \binom{j-1}{0} + \binom{j-1}{1} + \cdots + \binom{j-1}{k}$ additions, and the posterior using about $\binom{0}{k}_* + \binom{1}{k}_* + \cdots + \binom{n-1}{k}_* = \binom{n}{k+1}_* - 1 > \binom{n}{k+1}$ additions (followed by $n-1$ multiplications). This is tolerable for small values of $n$ and $k$, but for larger values, say $n = 60$ and $k = 5$, the time requirement becomes infeasible in practice, for the computations are performed for thousands of node orderings.

The same concern holds for the possible second phase of the order-MCMC method, in which each sampled node ordering is used for sampling some number of DAGs compatible with the ordering. This amounts to $n$ subset sampling queries per DAG.

### 1.2 AN APPROXIMATION APPROACH

One might hope for a data structure that enables rapid answering of subset queries. Trivially, counting queries can be answered in $O(n)$ time by precomputing and storing the answers to all possible $2^n$ queries in advance. But this approach becomes soon unfeasible for larger $n$, especially due to the large memory requirement. It is to be contrasted with the other extreme approach: visit all members $S \in \mathcal{C}$ and add $w(S)$ to the sum if $S \subseteq Q$, taking $O(mn)$ time and essentially no extra memory. Whether there are efficient ways to trade memory for time, is an open question. However, there is strong negative evidence associated with closely related existence queries ("Does the query set contain some set in $\mathcal{C}$?"). Namely, the best known tradeoffs are rather inefficient and of theoretical interest only (Charikar et al., 2002), and lower bounds that suggest the impossibility of finding much better tradeoffs are known (Pătraşcu, 2011).

Given the state of affairs concerning exact solutions, we in this work settle for approximations to reduce the time requirement of subset queries. In counting queries we allow an additive relative error $d$. Likewise, in sampling queries we require the sampling distribution be at a total variation distance at most $d$ from the exact distribution. Here $d$ is a parameter that can be set to close to zero, say $d = 0.01$, to guarantee very accurate approximations. In the order-MCMC application, for instance, it is easy to verify that accurate approximations to subset queries translate, in a straightforward manner, to accurate approximations at the end results of posterior inference.

The idea of approximation is not new. Indeed, the starting point of the present study is the following heuristic by Friedman and Koller (2003) to speed up order-MCMC: Given a query set $Q$, visit a fixed number $m'$ of heaviest members $S \in \mathcal{C}$ in decreasing order by weight, and if $S \subseteq Q$, then increase the sum by the weight of $S$. Finally, return the accumulated sum, unless the largest counted weight fails to be some factor $\gamma$ larger than the smallest (last) weight in the list, in which case compute the sum by brute-force enumeration of all subsets of $Q$ in $\mathcal{C}$. The rationale is that for large $Q$ it is likely that there is at least one heavy $S \subseteq Q$ among the $m'$ heaviest sets in $\mathcal{C}$, and thus the brute-force phase is avoided. Choosing a large enough $\gamma$ guarantees that the lost mass is negligible. Whether this heuristic is close to the best possible, has remained an open question.

Here, we address the question in several ways. Our focus is exclusively on *collector* algorithms that, like the aforementioned heuristic by Friedman and Koller, are based on visiting some of the subsets of the ground set in some order, adding up the weights of those that are subsets of the query set, and stopping using some appropriate rule. We begin in Section 2 by showing how any algorithm of this type for approximative counting also yields a sampling algorithm with a corresponding accuracy. The section continues by describing two extreme approaches to counting queries: a brute-force algorithm *Exact* that produces the exact value, and an idealized algorithm *Ideal* that only visits the minimum number of heaviest subsets that suffice for the desired approximation error, so providing us a lower bound for the amount of work needed by any collector algorithm. We also streamline the heuristic of Friedman and Koller by formulating a stronger stopping rule that achieves the same accuracy guarantees with less work. We call the resulting algorithm *Sorted*.

Our main contribution is a novel heuristic, presented in Section 3. The motivation of the heuristic stems from the observation that *Sorted* becomes slow when the heaviest subsets are not contained by the query set. Our idea is to restrict the search to subsets of the query set, however, turning the brute-force enumeration into a controlled approximation algorithm that, in a greedy fashion, aims to visit first subsets that are "likely" to be heavier. As the enumeration proceeds from smaller subsets to larger ones in a tree-structured manner, we call the algorithm *Treedy*.

We compare the heuristics to the exact and the idealized algorithm in Section 4. We report on experiments with synthetic instances and in application to Bayesian network learning using order-MCMC.

To keep the presentation simple and succinct, we will assume that the collection $\mathcal{C}$ is *downward closed*. That is, we assume that $\mathcal{C}$ equals the *downward closure* $\mathcal{C}_* = \{T : T \subseteq S \text{ for some } S \in \mathcal{C}\}$. In the experiments we further restrict our attention to the case where $\mathcal{C}$ consists of all $S \subseteq N$ with $|S| \leq k$ for some relatively small $k$. These restrictions are, however, not crucial for the validity of the studied methods. We discuss this issue, among other things, in Section 5.

## 2 PRELIMINARIES

Throughout this section and the remainder of the paper, we consider a weighted downward closed collection $\mathcal{C}$ of $m$ subsets of some $n$-element ground set $N$. For a query set $Q \subseteq N$, we call a set $S$ *relevant* if $S \in \mathcal{C}$ and $S \subseteq Q$. We denote the collection of relevant sets by $\mathcal{C}_Q$. If $Q$ is clear from the context, we may denote the total weight $W(Q)$ simply by $W$. We will denote by $d$ the approximation *tolerance*, $0 \leq d \leq 1$, whether referring to an upper bound for the additive relative error or for the total variation distance.

### 2.1 FROM COUNTING TO SAMPLING

Below is a generic algorithm that uses a collector algorithm for counting queries (the first step) to answer sampling queries (the second step).

**Algorithm *Draw***
Given a query set $Q$ and tolerance $d$, do the following:

**D1** Visit some relevant sets $S_1, S_2, \ldots, S_r$ whose total weight $W'$ is at least $(1-d)W$. Store the cumulative sums $W_i = \sum_{j=1}^{i} w(S_j)$.

**D2** Draw a random variable $U$ from the uniform distribution on the interval $[0, W']$. Find an $i$ such that $W_{i-1} < U \leq W_i$ and return $S_i$.

It is easy to see that *Draw* returns a set $S$ with probability $\pi'(S)$ that satisfies $\pi'(S) = w(S)/W'$ if $S$ is a visited relevant set and $\pi'(S) = 0$ otherwise. The next result shows that this guarantees a small deviation from the exact distribution $\pi(S) = w(S)/W$, as measured by the commonly used total variation distance. The *total variation distance* between two probability measures $\mu$ and $\mu'$ on a finite set $\Omega$ is defined as $\delta(\mu, \mu') = \max_{A \subseteq \Omega} |\mu(A) - \mu'(A)|$ and can be simplified to $\delta(\mu, \mu') = \sum_{a \in \Omega} |\mu(a) - \mu'(a)|/2$. We attribute the following theorem to folklore; the proof is elementary and included here for convenience.

**Theorem 1.** *The total variation distance between the above defined $\pi$ and $\pi'$ is at most $d$.*

*Proof.* As $\pi'(S) \geq \pi(S)$ for each visited relevant set $S$ and $\pi'(S) = 0$ otherwise, we have

$$
\begin{aligned}
2\delta(\pi, \pi') &= \sum_{i=1}^{r} \left[\pi'(S_i) - \pi(S_i)\right] + 1 - \sum_{i=1}^{r} \pi(S_i) \\
&= 2 - 2W'/W\,.
\end{aligned}
$$

Using $W' \geq (1-d)W$ completes the proof. □

If the cumulative weights $W_i$ are stored in a simple array indexed by $i$, step D2 can be implemented to run in $O(\log r)$ time by using binary search. However, if the distribution $\pi'$ has small entropy $H$ (i.e., the mass is concentrated on some subsets), then much faster implementation running in roughly $O(H)$ time is possible by using, e.g., a Huffman coding based data structure.

The running time of *Draw* is clearly dominated by step D1. This, in part, motivates the investigations of efficient collector algorithms for counting queries.

### 2.2 EXACT AND IDEAL COUNTING

There are two obvious brute-force approaches to compute the exact total weight of the query set $Q$. One is to visit every set $S$ in the given collection $\mathcal{C}$ and add the weight of $S$ to a cumulative sum if $S \subseteq Q$. The other approach is to only visit sets $S \subseteq Q$ and add the weight of $S$ to a cumulative sum if $S \in \mathcal{C}$. The following algorithm assumes the latter approach:

**Algorithm *Exact***
Given a query set $Q$, do the following:

**E1** Visit the relevant sets in lexicographic order and return the sum of their weights.

For downward closed collections $\mathcal{C}$ it is, in fact, more efficient to implement the algorithm so that the membership test $S \in \mathcal{C}$ is avoided, at the cost of visiting also a few sets that are not subsets of $Q$. The idea is to use a data structure where each member $S$ of $\mathcal{C}$ is linked to its one-element larger successors $S \cup \{x\} \in \mathcal{C}$, with the link labeled by the element $x$. Namely, then the algorithm can proceed in the lexicographic order at the cost of testing whether $x \in Q$ also for some some irrelevant sets $S \cup \{x\} \not\subseteq Q$. Technically, this makes the number of visited sets generally exceed the number of relevant sets $|\mathcal{C}_Q|$; however, the extra visits are very quick due to the simplicity of the test.

We note that when $\mathcal{C}$ consists of all subsets of size at most some $k$, then it is easy to visit only the relevant sets and avoid the aforementioned technicalities.

When an approximation of the total weight suffices, the performance of *Exact* is no longer close to the best possible. An ideal collector algorithm would visit as few as possible sets necessary for gathering the required proportion of the mass:

**Algorithm *Ideal***
Given a query set $Q$ and tolerance $d$, do the following:

**I1** Visit, in some order, the minimum number of the heaviest relevant sets whose total weight $W'$ is at least $(1-d)W$. Return $W'$.

We should note that *Ideal* is an "idealized" algorithm in the sense that we do not know how to efficiently find the minimum number of the heaviest relevant sets. Nevertheless, we can estimate the running time of the algorithm under the supposition that a list of sufficient relevant sets is available to the algorithm for free.

### 2.3 COUNTING BY PRE-SORTING

If the query set $Q$ is large, then one can achieve almost ideal performance by visiting sufficiently many members of $\mathcal{C}$ in decreasing order by weight. Sorting needs to be done only in the initialization phase, before any query. For smaller $Q$ the heaviest sets of $\mathcal{C}$ are, however, likely to include also irrelevant sets that are not subsets of $Q$. Then the number of visited sets may grow much larger than $|\mathcal{C}_Q|$. Therefore it is advisable to switch over to the exact algorithm after about $|\mathcal{C}_Q|$ visited sets. To keep the number of visited sets as small as possible, it is also crucial to devise an efficient stopping rule. The following algorithm adapts these ideas and implements a stopping condition that is stronger than in the original formulation by Friedman and Koller (2003):

**Algorithm *Sorted***

**S0** Before any query, sort the sets in $\mathcal{C}$ into decreasing order by weight, $w(S_1) \geq w(S_2) \geq \cdots \geq w(S_m)$. Store the cumulative sums $W_i = \sum_{j=i+1}^{m} w(S_j)$.

**S1** Given a query set $Q$ and tolerance $d$, initialize a counter for yet nonvisited relevant sets $t = |\mathcal{C}_Q|$, a step counter $j = 0$, and $W' = 0$, and do the following: increase $j$ by 1; if $j \geq |\mathcal{C}_Q|$, then switch to *Exact*; if $S_j \subseteq Q$, then add $w(S_j)$ to $W'$ and decrease $t$ by 1; if $W' \geq (1-d)(W' + W_j - W_{j+t})$, then stop and return $W'$.

**Theorem 2.** *Algorithm Sorted is correct.*

*Proof.* If the algorithm switches to *Exact* at some point, then the correctness of the algorithm is clear. Suppose therefore that the algorithm stops when the condition $W' \geq (1-d)(W' + W_j - W_{j+t})$ is satisfied. It suffices to show that $W' + W_j - W_{j+t} \geq W$. To this end, observe that $W'$ is a sum of the weights of the $|\mathcal{C}_Q| - t$ heaviest relevant sets, and that $W_j - W_{j+t}$ is at least as large as the sum of the $t$ remaining (lightest) relevant sets. □

Table 1: Processing a Subset Counting Query. For each algorithm the visiting order of subsets is shown, for ground set {A, B, C, D}, query set {B, C, D}, and approximation tolerance 20 %.

| $S \in \mathcal{C}$ | $w(S)$ | Exact | Ideal | Sorted | Treedy |
|---|---|---|---|---|---|
| AB | 99 | | | 1 | |
| AD | 90 | | | 2 | |
| A | 85 | 2 | | 3 | 2 |
| ∅ | 80 | 1 | ✓ | 4 | 1 |
| B | 70 | 3 | ✓ | 5 | 3 |
| AC | 60 | | | 6 | |
| D | 50 | 8 | ✓ | 7 | 4 |
| BD | 14 | 5 | | | 5 |
| C | 13 | 6 | | | 6 |
| CD | 12 | 7 | | | |
| BC | 11 | 4 | | | |

See Table 1 for an illustration and comparison to *Exact* and *Ideal*. Note that in the example, *Sorted* achieves the desired approximation guarantee after visiting 7 sets; it does not switch to *Exact*.

## 3 THE TREEDY HEURISTIC

We next present a novel heuristic, *Treedy*, for approximate counting queries. Like *Exact*, the heuristic operates on a lexicographically structured tree. The *lexicographical tree* of $\mathcal{C}$ is a rooted tree on the collection $\mathcal{C}$, defined as follows. The empty set is the root of the tree. Any other set $S$ is a *son* of another set $S'$ if $S = S' \cup \{x\}$ where $x$ is the last element in $S$ in alphabetical order; the notions of a *brother*, *ancestor*, and *descendant* are defined in the obvious way. For each set $S \in \mathcal{C}$ define the *weight potential* $\phi(S)$ as the sum of the weights of all descendants of $S$ (including $S$ itself). In the initialization phase, *Treedy* modifies this structure appropriately, and in the query phase it proceeds in a greedy fashion.

**Algorithm *Treedy***

**T0** Before any query, modify the lexicographical tree of $\mathcal{C}$ into a *greedy tree* as follows: For each set, sort its sons in decreasing order by their weight potentials. Then remove the links to all but the first son and create a linked list to connect the brothers in this order. As a result, each set has (at most) two links: one to its brother with next largest weight potential and the other to its son with the largest weight potential. Finally compute the *aggregate potential* $\psi(S)$ for each set $S$ by summing the weight potentials of $S$ and its subsequent smaller brothers.

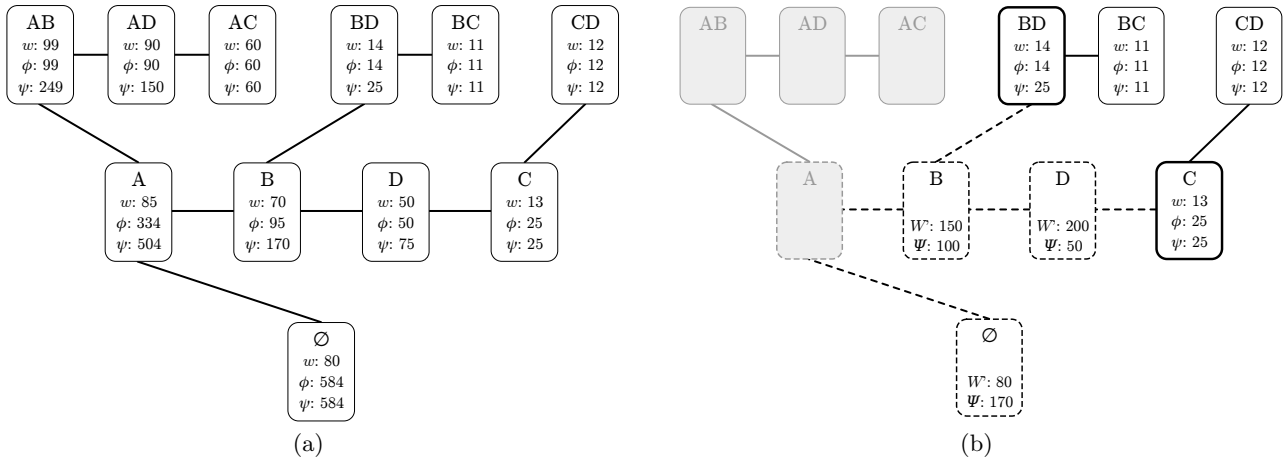

Figure 1: *Treedy* in action on the example instance described in Table 1. (a) The greedy tree built in step T0 is shown. (b) The situation at the end of step T1 is shown. Dashed lines connect the visited sets. The sets marked by thicker rectangles constitute the collection $\mathcal{R}$ at the end of step T1. The last set removed from $\mathcal{R}$ is $\{D\}$. A subtree that is discarded once found to contain only irrelevant sets is shown in gray. Note that $\phi$ and $\psi$ remain unchanged after step T0; only the values of $W'$ and $\Psi(\mathcal{R})$ change during the execution of step T1 (for each set shown are the values just after removing the set from $\mathcal{R}$ and updating $W'$).

**T1** Given a query set $Q$ and tolerance $d$, initialize $W' = 0$ and $\mathcal{R} = \{\emptyset\}$ and repeat the following: remove from $\mathcal{R}$ a set $S$ with the largest aggregate potential; add $w(S)$ to $W'$; add the next relevant brother (if any) of $S$ into $\mathcal{R}$ and ignore the preceding irrelevant brothers (and their descendants); add the first relevant son (if any) of $S$ into $\mathcal{R}$ and ignore the preceding irrelevant sons (and their descendants); if $W' \geq (1-d)\big(W' + \Psi(\mathcal{R})\big)$, where $\Psi(\mathcal{R}) = \sum_{S \in \mathcal{R}} \psi(S)$ is the aggregate potential of $\mathcal{R}$, then stop and return $W'$.

See Table 1 and Figure 1 for an illustration of an execution of *Treedy*. In that small example, the savings of *Treedy* compared to *Exact* are rather modest. We leave it to the reader to imagine how larger savings are possible on larger problem instances.

**Theorem 3.** *Algorithm Treedy is correct.*

*Proof.* First note that, during the execution of the algorithm, $\mathcal{R}$ contains only relevant sets and thus only weights of relevant sets are added to $W'$. Therefore, it is sufficient to show that the invariant $W' + \Psi(\mathcal{R}) \geq W(Q)$ holds during the execution of the algorithm; correctness then follows from the stopping condition.

To see that the invariant holds, observe that in the beginning $\Psi(\mathcal{R})$ is the sum of the weights of all relevant sets plus the weights of all irrelevant sets. In each step the algorithm removes a set $S$ from $\mathcal{R}$ but adds back the next relevant brother of $S$ and the first relevant son of $S$. The aggregate potential of $\mathcal{R}$ is thus decreased by $w(S)$ and by the weight potentials of any possibly ignored irrelevant brothers and sons. Since the descendants of irrelevant sets are also irrelevant, the only relevant set whose weight is subtracted from $\Psi(\mathcal{R})$ is $S$. As $w(S)$ is added to $W'$, no weight of a relevant set is removed from $W' + \Psi(\mathcal{R})$. Thus, by the induction principle the invariant holds during the execution of the algorithm. □

The potential efficiency of *Treedy* stems from the fact that, contrary to *Sorted*, none of the descendants of an irrelevant set will be visited. On the other hand, the algorithm is allowed to visit some irrelevant sets to enable efficient implementation of the greedy best-first order. But visiting irrelevant nodes is fast since those are just ignored; no update of $\mathcal{R}$, $\Psi(\mathcal{R})$ or $W'$ is needed.

The greedy tree can be constructed in a straightforward manner in $O(m \log n)$ time, for sorting the $m_S$ sons of each set $S \in \mathcal{C}$ takes $O(m_S \log m_S)$ time, $m_S \leq n$, and $\sum_{S \in \mathcal{C}} m_S = m - 1$. Thus the initialization cost is essentially linear in the input size and negligible when the number of counting queries is large.

To implement the query algorithm efficiently, we have to overcome two challenges: (1) how to store $\mathcal{R}$ and (2) how to avoid computing $\Psi(\mathcal{R})$ from scratch in every step. We address the first challenge by keeping the members of $\mathcal{R}$ in a binary heap, which enables updating $\mathcal{R}$ in $O(\log|\mathcal{R}|)$ time per step of the algorithm.

To address the second challenge, we maintain the sum $\Psi(\mathcal{R})$ during the execution of the algorithm by, in each step, subtracting from $\Psi(\mathcal{R})$ the aggregate potential of the set removed from $\mathcal{R}$ and adding to $\Psi(\mathcal{R})$ the aggregate potentials of the sets added to $\mathcal{R}$. However, if the relative differences of the weights $w(S)$ are large, then this approach can lead to problems with numerical accuracy. If this is the case, a solution is to not maintain $\Psi(\mathcal{R})$ at all until we know that the stopping condition is relatively close to hold. More specifically, suppose $\Psi(\mathcal{R})$ is computed from scratch and maintained only after the aggregate potential of the set removed from $\mathcal{R}$ gets smaller than $dW'$. Then we know that $\Psi(\mathcal{R})$ has to decrease at most by the factor $|\mathcal{R}|$ before the stopping condition is met, and thus the accuracy problems should be gone.

## 4  EXPERIMENTAL STUDIES

We have implemented the presented algorithms in the C++ language.[1] We next report on experimental studies on artificially generated instances of subset query problems as well as several instances of the Bayesian network learning application using our implementation of the order-MCMC method of Friedman and Koller (2003). The focus of the experimental studies is in investigating the relationship of running time and approximation guarantee of the four algorithms.

### 4.1  ARTIFICIAL INSTANCES

We generated various weight functions $w(S)$ on subsets $S \subseteq N$ of size at most $k$. We varied $n = |N|$ in $\{20, 60\}$ and $k$ in $\{3, 5\}$. We considered four types of weight functions, each based on the following building block:

$$w(S) = \exp\left(\lambda \sum_{i \in S} U_i\right), \quad U_i \stackrel{\text{iid}}{\sim} \text{Uniform}(\kappa - 1, \kappa).$$

Here $\lambda$ is a parameter that specifies the variance of the weights; the larger the $\lambda$, the larger the variance. The parameter $\kappa$ specifies the (expected) number of elements of the ground set that contribute positively to the weight. Note that the weight of the empty set is always 1. The four types are the following:

**Flat:** $\lambda = 10$, $\kappa = k/n$.

---
[1]The implementation will be made publicly available via the authors' home pages.

**Steep:** $\lambda = 200$, $\kappa = k/n$.

**Mixture:** Take the sum of 5 flat and 5 steep ones, for both types letting the product $\kappa n$ take the 5 values $k-1$, $k$, $k+1$, $k+2$, and $k+3$. This creates 10 distinct "local maxima".

**Shuffled:** Like in the mixture type but permuting the weights $w(S)$ randomly among the subsets of same size. This destroys the dependence of the weights of subsets that have large intersection.

For each of the four types and the values of $n$ and $k$, we generated 5 random weight functions, executed the algorithms *Exact*, *Ideal*, *Sorted*, and *Treedy*, for 1000 query sets sampled uniformly at random for each query set size from 1 to $n$. Figure 2 shows average running times.

We see that for flat weight functions, the approximation algorithms yield significant speedups over the exact algorithm only when the approximation tolerance is relatively large. For steep weight functions, as well as for mixtures and shuffled mixtures, the speedups are however by two orders or magnitude already with small approximation tolerance. The speedups become the more dramatic, the larger the $n$ and $k$ are. Examining the effect of the query set size reveals that *Sorted* performs better than *Treedy* for larger query sets, but for smaller query sets *Treedy* is faster. For larger values of $n$ and $k$, *Treedy* outperforms *Sorted* by a factor of about 5. To our surprise, shuffling the weight function has essentially no effect to the performance of *Treedy*.

### 4.2  APPLICATION TO BAYESIAN NETWORK LEARNING

We ran our implementation of order-MCMC on four datasets available from the UCI repository (Blake and Merz, 1998). Votes2 was obtained by concatenating two random permutations of the 17-variable Votes dataset, so doubling the number variables. 10xPromoters was obtained by taking each sample of the Promoters dataset 10 times. The datasets Chess and Splice we used as such. In addition, we used datasets of varying sample sizes generated from the benchmark Bayesian network models Alarm (Beinlich et al., 1989) and Hailfinder (Abramson et al., 1996).[2] Table 2 shows the key parameters associated with the datasets and the order-MCMC method, including the maximum number of parents $k$ and the number of MCMC steps. The posterior (the structure and parameter priors) was specified as in the experiments of Niinimäki et al. (2011).

Table 2: Datasets, Generative Bayesian Network Models, and Parameter Settings Used in the Experiments for Bayesian Network Learning by Order-MCMC.

| Name | $n$ | #Samples | $k$ | #Steps |
|---|---|---|---|---|
| Votes2 | 34 | 435 | 5 | 10000 |
| Chess | 37 | 3196 | 5 | 5000 |
| 10xPromoters | 58 | 1060 | 4 | 2000 |
| Splice | 61 | 3190 | 4 | 2000 |
| | | | | |
| Alarm | 37 | 50–5000 | 5 | 1000 |
| Hailfinder | 56 | 50–5000 | 4 | 1000 |

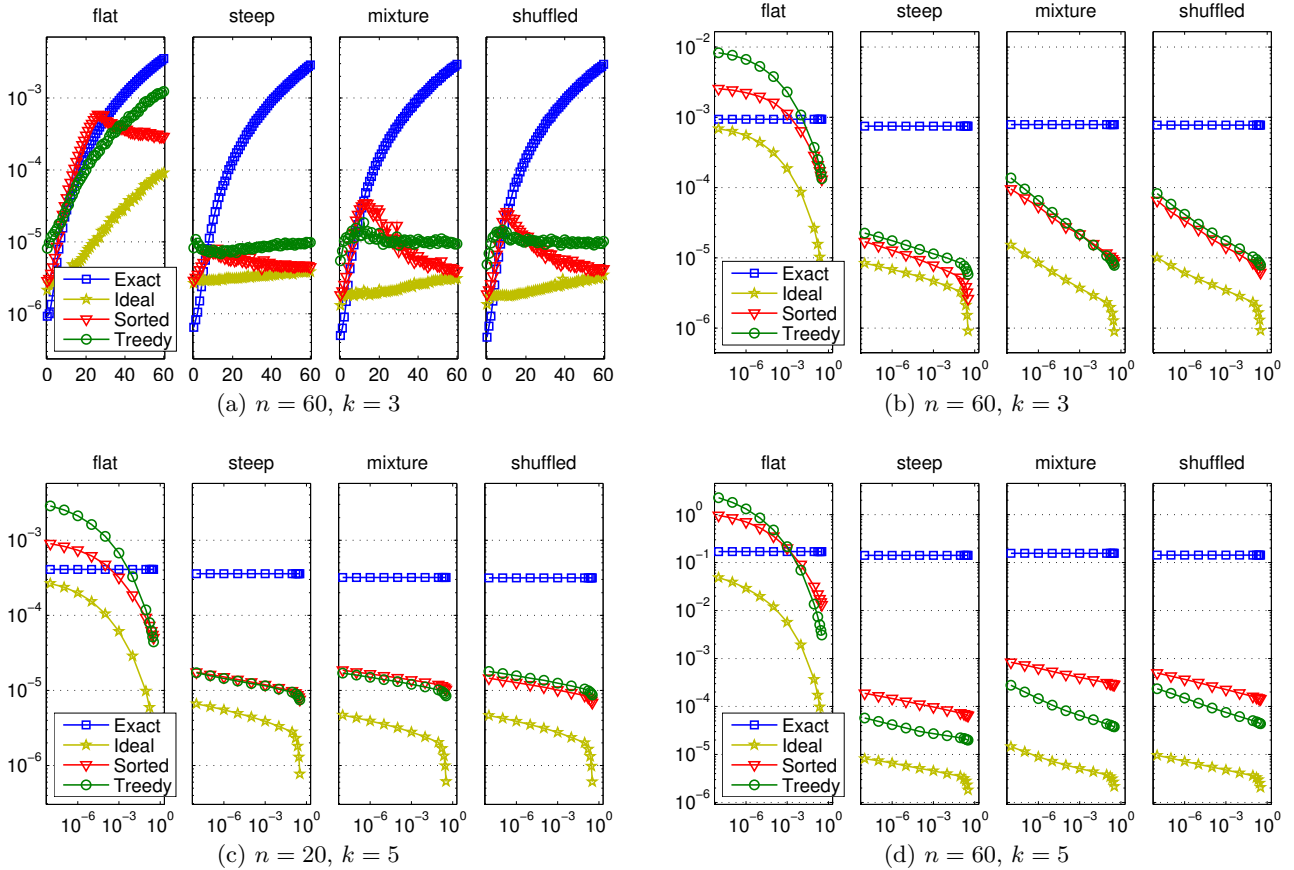

Figure 2: Runtime in artificial problem instances. The number of seconds per subset counting query for *Exact*, *Ideal*, *Sorted*, and *Treedy* are shown (a) as a function of the query set size, for a fixed approximation tolerance of 0.1, and (b, c, d) as a function of approximation tolerance.

We observe that approximation expedites the computations by one to several orders of magnitude; see Figure 3. As expected, the gain of approximation increases with larger datasets and larger error, being, however, significant already with 200 samples and easily tolerable error (say 1%). On the larger datasets *Treedy* performs consistently better than *Sorted*, the difference being sometimes nearly one order of magnitude (Chess and Alarm with an approximation tolerance of at least 1%).

Examining the effect of the query set size for the four datasets (Figure 3(a)) reveals that *Treedy* is consistently faster than *Sorted* on queries that are the hardest ones for *Sorted*. However, *Sorted* is typically faster than *Treedy* on the easier query sets. Thus it depends on the distribution of queries, whether *Treedy* or *Sorted* should be the algorithm of choice, or whether it would pay off to use the obvious hybrid: *Treedy* for smaller and *Sorted* for larger query sets.

The results for *Ideal* suggest that considerable further speedups, by one to two orders of magnitude, might be possible using still better algorithms and data structures.

---

[2]The datasets are available at http://www.dsl-lab.org/supplements/mmhc_paper/mmhc_index.html.

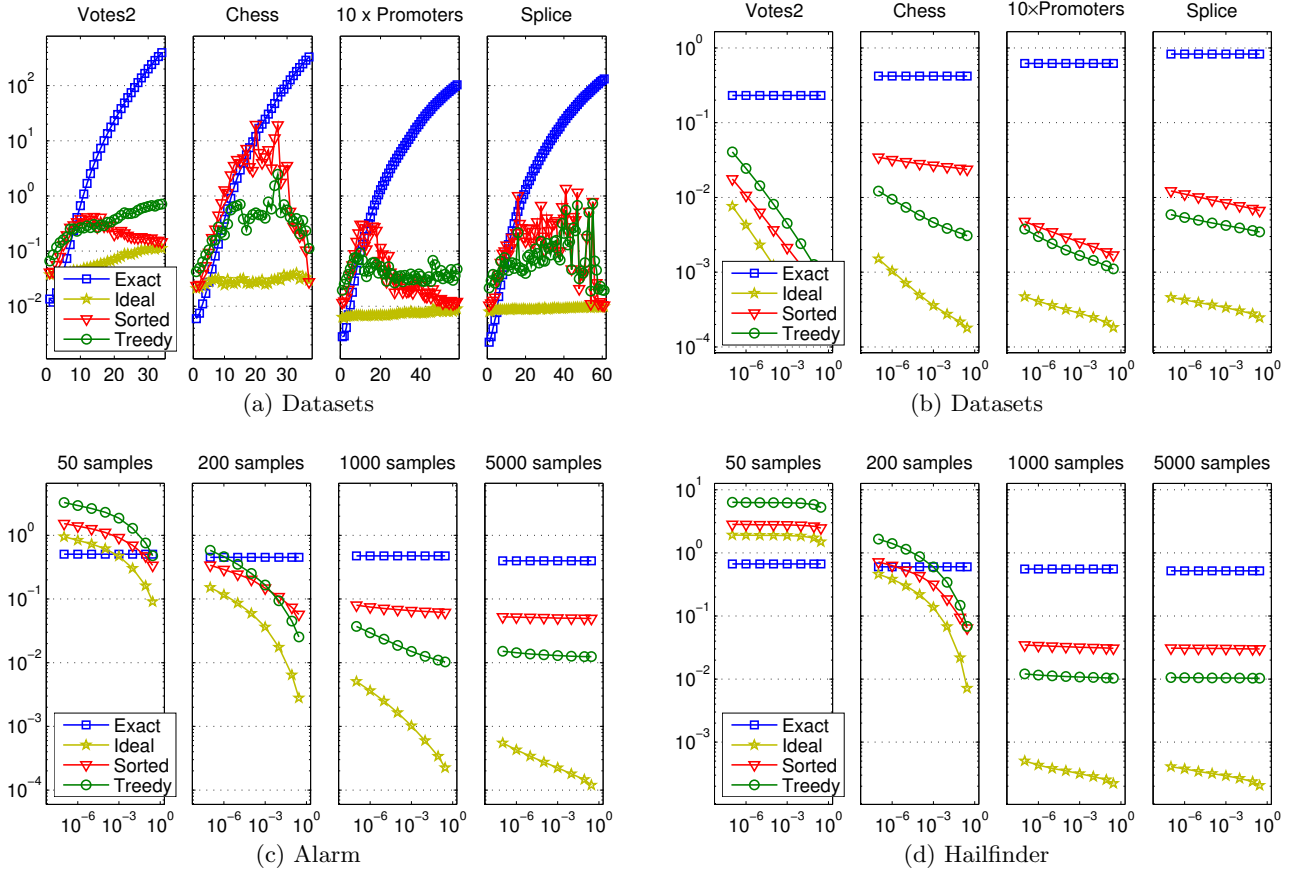

Figure 3: Runtime in the Bayesian network application. The number of seconds per MCMC step for *Exact*, *Ideal*, *Sorted*, and *Treedy* are shown (a) as a function of the query set size, for a fixed approximation tolerance of 0.1, and (b, c, d) as a function of approximation tolerance $d$. Since the approximated posterior probability is obtained as a product of $n$ approximated total weights, the tolerance per total weight was set to $d/n$ to guarantee the required accuracy.

## 5 DISCUSSION

We have studied approximation algorithms for subset counting and sampling queries. After observing how any collector algorithm for approximate counting queries can be turned into a sampling algorithm, we considered four collector algorithms. These include a slow exact algorithm that serves as a reference and an "ideal" one that we cannot implement efficiently but that provides us with a lower bound of the needed work. Our experimental results suggest that the two heuristic methods, *Sorted* and *Treedy*, can yield dramatic speed-ups (by several orders of magnitude) over the exact algorithm, while not quite achieving the ideal performance. Typically, *Treedy* performs as well as *Sorted* or significantly better.

We made the assumption that the given collection of subsets is downward closed. This assumption simplified the presentation and experimental settings. The assumption is, however, not well justified in general. Namely, the input collection can potentially be much smaller than its downward closure, in which case one could realistically hope for faster methods whose time requirement is determined by the size of the input rather than the size of the closure. We note that *Sorted* readily has this desirable property. For example, in the Bayesian network application it is quite plausible to expect that some potential parent sets have so small a weight that they can be discarded in the precomputation phase. Fortunately, the assumption of downward closedness seems not crucial for the validity of the presented methods. Indeed, the data structures and visiting orders underlying the methods *Exact* and *Treedy* can be pruned by introducing shortcuts. Using shortcuts, only subsets that belong to the collection need be visited, and so the other subsets can be discarded. We leave a more detailed description of this generalization and examination of its impact to the Bayesian network application to an extended version of this paper.

There are also other avenues for future research. We restricted our attention to methods whose memory requirement is roughly linear in the size of the input collection. It remains an open question, whether significant speedups can be achieved by investing somewhat more space. Likewise, we have only considered collector algorithms, and it is an open question, whether there exist faster algorithms of some different type.

**Acknowledgements**

The authors thank Petteri Kaski, Janne Korhonen, and Pekka Parviainen for valuable discussions about various aspects of the zeta transform, and the anonymous reviewers for useful comments on earlier versions of this paper. This research was supported in part by the Academy of Finland, Grants 125637, 218153, and 255675.